\documentclass{aa}

\usepackage{graphicx}
\usepackage{txfonts}

\newcommand{\Mpc}{$h^{-1}$\thinspace Mpc}

\newcommand{\etal}{et al.}
\def\apj{ApJ}
\def\apjl{ApJL}
\def\apjs{ApJS}

\def\aap{A\&A}

\begin{document}

\title{Toward Understanding Environmental Effects in SDSS Clusters}

\author{J. Einasto\inst{1}, E. Tago\inst{1}, M.  Einasto\inst{1},
  E. Saar\inst{1}, I. Suhhonenko\inst{1}, 
  P. Hein\"am\"aki\inst{1,2} G. H\"utsi\inst{1,3}\&  D. L. Tucker\inst{4}}

\institute{Tartu Observatory, EE-61602 T\~oravere, Estonia 
\and 
Tuorla Observatory, V\"ais\"al\"antie 20, Piikki\"o, Finland 
\and 
Max-Planck Institut f\"ur Astrophysik, Karl-Schwarzschild-Str 1, 86740
Garching, FRG
\and
 Fermi National Accelerator Laboratory, MS 127, PO Box 500, Batavia,
IL 60510, USA } 

\date{ Received  09.11.2004, Accepted
} 

\authorrunning{J. Einasto et al.}
\titlerunning{Environmental Effects in SDSS clusters}

\offprints{J. Einasto }

\abstract { We find clusters and superclusters of galaxies using the
Data Release 1 of the Sloan Digital Sky Survey. We calculate a
low-resolution density field with a smoothing length of 10~\Mpc\ to
characterise the density of the cluster environment.  We determine the
luminosity function of clusters, and investigate properties of
clusters in various environments.  We find that clusters in a
high-density environment have a luminosity a factor of $\sim 5$ higher
than in a low-density environment.  We also study clusters and
superclusters in numerical simulations.  Simulated clusters in a
high-density environment are also more massive than those in a
low-density environment.  Comparison of the density distribution at
various epochs in simulations shows that in large low-density regions
(voids) dynamical evolution is very slow and stops at an early epoch.
In contrast, in large regions of higher density (superclusters)
dynamical evolution starts early and continues until the present; here
particles cluster early, and by merging of smaller groups very rich
systems of galaxies form.

\keywords{cosmology: large-scale structure of the Universe -- clusters
of galaxies; cosmology: large-scale structure of the Universe --
Galaxies; clusters: general; cosmology: simulations; cosmology: evolution}

}

\maketitle

\section{Introduction}

Clusters and superclusters of galaxies are the basic building blocks
of the Universe on cosmological scales.  The first catalogues of
clusters of galaxies by Abell (\cite{abell},~\cite{aco}) and Zwicky et
al. (\cite{zwicky}) were constructed by visual inspection of the
Palomar Observatory Sky Survey plates.  Modern surveys of galaxies,
such as the Las Campanas Redshift Survey (LCRS), the Sloan Digital Sky
Survey (SDSS) and the two-degree-field (2dF) Galaxy Redshift Survey,
enable us to define groups, clusters and superclusters of galaxies and
to investigate properties of these galaxy systems in various
large-scale environments.  

Studies of the dependence of properties of
galaxy systems on the density of the large-scale environment have been
made by Einasto et al. \cite{e03a}, \cite{e03b}, \cite{e03c},
\cite{e03d} (hereafter E03a, E03b, E03c and E03d, respectively), using
the Early Data Release (Stoughton et al.  \cite{s02})of the SDSS and
the Las Campanas Redshift Survey.  These studies demonstrated the
presence of environmental effects in clusters -- clusters in a
high-density environment are richer and larger than in a low-density
environment.

The present study has three goals. First of all, we shall check
the results obtained by E03a and E03b using a more accurate definition of
groups and clusters.  E03a and E03b found clusters of galaxies as
density enhancements in the high-resolution 2-dimensional density
field.  Such a definition has its restrictions, as in some cases clusters
may overlap in the projection direction, and cluster properties may
be distorted.  In contrast to previous studies we shall now define
groups and clusters in the conventional way using the full 3-dimensional
data on the distribution of galaxies. In this analysis we shall use
galaxy samples of the Data Release 1 of the Sloan Digital Sky Survey
(DR1 of SDSS, Abazajian et al. \cite{abazajian03}), and shall investigate
properties of these systems in relation to the large-scale
environment, from rich superclusters to poor filaments of loose groups
in voids.

The second goal of our study is to compare properties of clusters and
superclusters with properties of similar systems found in N-body
simulations of structure evolution.  In particular, we are interested
in the relationship of cluster properties and their large-scale
environment.  We define DM-haloes in simulations in the same way as
groups and clusters were defined in real galaxy samples, and
superclusters as large over-density regions of the smoothed density
field.  This comparison of real and
simulated cluster properties complements a similar study by Einasto et
al. (\cite{e04b}) where a different method was used to characterise
the density of the environment.

The third and ultimate goal of the present study is to try to find
an explanation for the environmental dependence of cluster
luminosities. For this purpose we shall compare the evolution of
groups and clusters in high- and low-density regions.  Also we shall
compare the distribution of particles located in systems of various
richness in high- and low-density environments.  This comparison shall
be done for various epochs, which gives us the possibility to follow
the evolution in regions of various global density.

In the next Section we describe the SDSS DR1 sample of galaxies and
the method used to find groups/clusters of galaxies.  Here we describe
also the N-body simulations of the structure evolution used to compare
observations with models. In Section 3 we describe the density field
of the SDSS DR1, and properties of clusters of the SDSS DR1. In
Section 4 we compare the properties of observed clusters with those of
similar objects found in simulations. In Section 5 we follow the
evolution of high- and low-density regions in an attempt to understand
the mechanism behind the environmental effects in cluster and galaxy
luminosities.  We discuss our results and compare them with previous
studies in Section 6.  The last Section brings our conclusions.
High-resolution colored figures of the SDSS DR1 density fields are
available at the web-site of Tartu Observatory ({\tt
http://www.aai.ee/$\sim$einasto}).  Preliminary results of this study
were reported at the conference on the Zone of Avoidance by Einasto et
al. (\cite{e04a}).

\section{Data}

\subsection{SDSS DR1 data}

The SDSS Data Release 1 consists of two slices of about 2.5 degrees
thick and 65--105 degrees wide, centered on the celestial equator, and
of several regions at higher declinations.  In the present study we
have used only the equatorial slices.  We extracted the Northern and
Southern slice samples from the full DR1 sample, using the following
criteria: the redshift interval $1000 \leq cz \leq 60000$ km~s$^{-1}$,
the Petrosian $r^*$-magnitude interval $13.0 \leq r^* \leq 17.7$, the
right ascension and declination intervals $145^{\circ} \leq RA \leq
250.0^{\circ}$ and $-1.25^{\circ} \leq DEC \leq 1.25^{\circ}$ for the
Northern slice, and $350^{\circ} \leq RA \leq 55.0^{\circ}$ and
$-1.25^{\circ} \leq DEC \leq 1.25^{\circ}$ for the Southern slice.
The number of galaxies extracted ($N_{\rm gal}$) and the length of the
slice in the right ascension ($\Delta RA$) are given in
Table~\ref{Tab1}.

{\scriptsize
\begin{table*}
      \caption[]{Data on the SDSS DR1 galaxies, clusters and superclusters}

         \label{Tab1}
      \[
         \begin{tabular}{ccccccccccccc}
            \hline
            \noalign{\smallskip}
            Sample & $\Delta$RA &
      	$\alpha_{E1}$&$M_{E1}^{\ast}$&  $\alpha_{E2}$&$M_{E2}^{\ast}$&
	    $\alpha_B$&$M_B^{\ast}$&
	$N_{\rm gal}$&$N_{\rm cl}$&$N_{\rm isol}$& $N_{\rm scl}$\\ 

            \noalign{\smallskip}
            \hline
            \noalign{\smallskip}

SDSS.N& $105^{\circ}$ &
$-1.06$&$-21.55$& $-1.22$ & $-20.80$& $-1.05$&$-20.44$& 19783&
2754&10232&26 \\ 

SDSS.S& $66^{\circ}$& $-1.10$ & $-20.71$ &
$-1.06$&$-21.40$&  $-1.05$&$-20.44$&11562&1451&6202&16\\ 
\\
            \noalign{\smallskip}
            \hline
         \end{tabular}
      \]
   \end{table*}
}

\begin{figure*}[ht]
\centering
\resizebox{0.45\textwidth}{!}{\includegraphics*{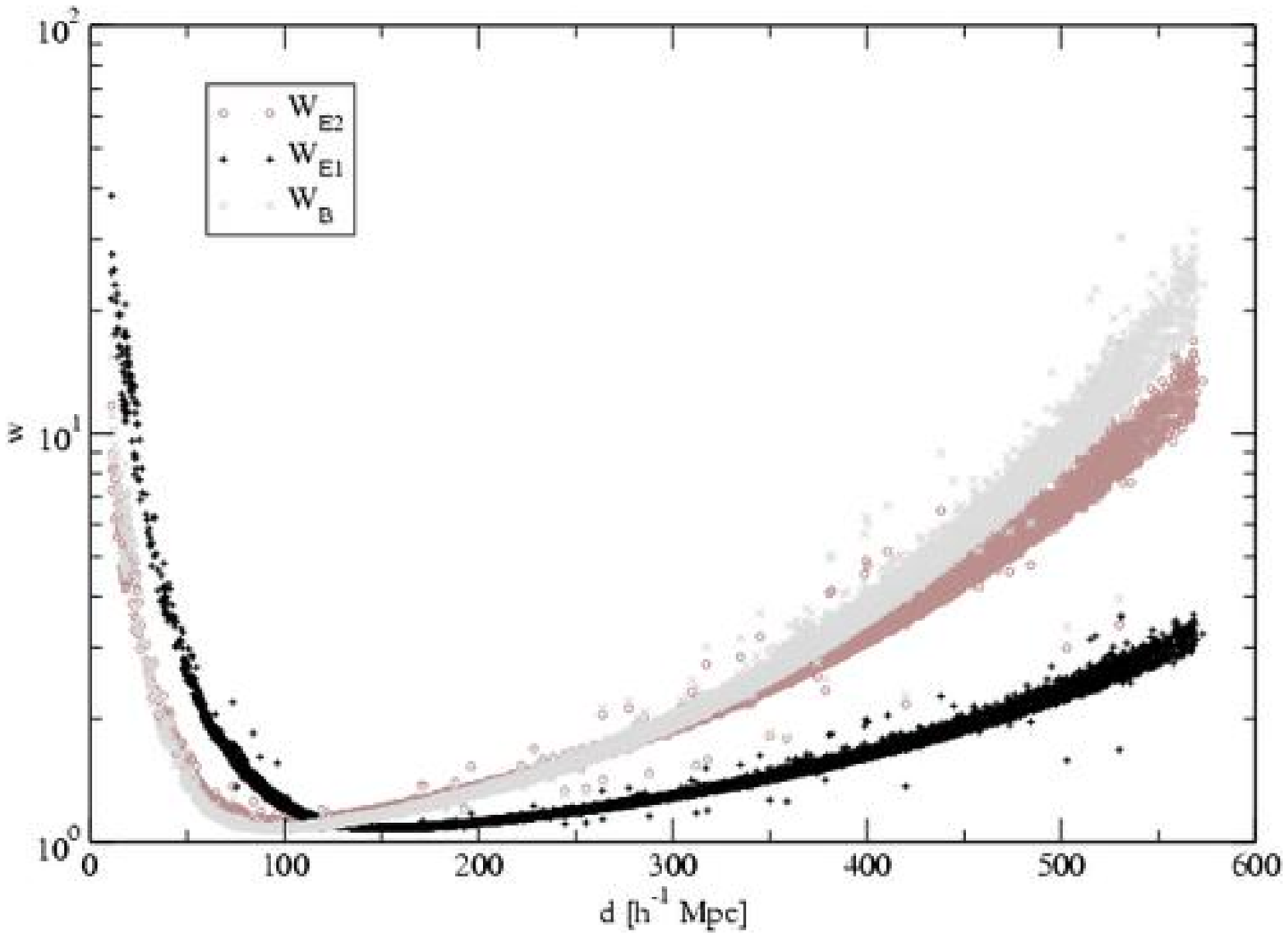}}
\hspace{2mm}
\resizebox{0.45\textwidth}{!}{\includegraphics*{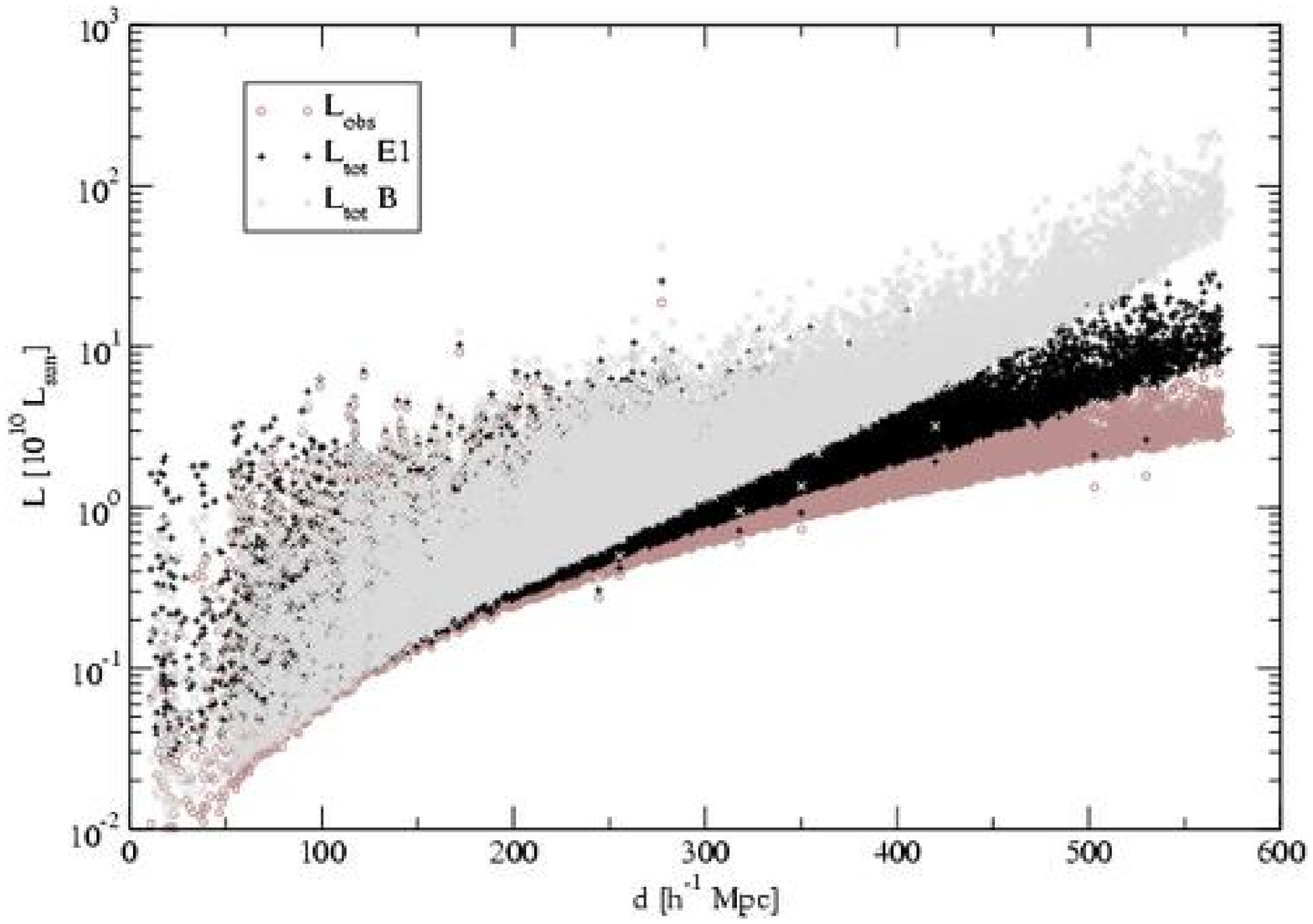}}
\hspace{2mm} 
\caption{The left panel shows the weights of galaxies used to correct
  for invisible galaxies outside the observational window. Black
  symbols show the weights for the Einasto set E1 of the Schechter
  parameters (used to find total luminosities of clusters); light gray
  and dark gray symbols show the weights for the Blanton B and the
  Einasto set E2 of the Schechter function parameters, respectively
  (the weights calculated with the set E2 were used to find the
  low-resolution density field). In the right panel we plot
  luminosities of galaxies: dark gray symbols show the observed
  luminosities, light gray and black symbols show the expected total
  luminosities, obtained using the Blanton B and the Einasto E1 sets
  of the Schechter function parameters.  }
\label{fig:1}
\end{figure*}

{\scriptsize
\begin{table*}
      \caption[]{Data on N-body DM-haloes and superclusters}

         \label{Tab2}
      \[
         \begin{tabular}{ccccccccccr}
            \hline
            \noalign{\smallskip}
            Sample & $L~ [Mpc/h]$ & $\Omega_m$ & $\Omega_{\Lambda}$ &
            $\sigma_8$ & $h$ & $N_p$&
	    $M_p[10^{10}h^{-1}M_\odot]$&
      	    $N_{min}$ & $N_{\rm halo}$ \\ 

            \noalign{\smallskip}
            \hline
            \noalign{\smallskip}

M100&  100 & 0.30 & 0.70 & 0.80 & 0.70 & $128^3$ & 3.98 & 8  & 2459  \\ 
M200A& 200 & 0.30 & 0.70 & 0.80 & 0.70 & $256^3$ & 3.98 &    &     \\ 
M200B& 200 & 0.27 & 0.73 & 0.84 & 0.71 & $256^3$ & 3.59 & 20 & 12306  \\ 

            \noalign{\smallskip}
            \hline
         \end{tabular}
      \]
   \end{table*}
}

The SDSS data reduction procedure consists of several steps: (1)
calculation of the distance, the absolute magnitude, and the weight
factor for each galaxy of the sample; (2) finding groups/clusters of
galaxies using the friends-of-friends algorithm; (3) calculation of
the density field using an appropriate kernel and a chosen smoothing
length. When calculating luminosities of galaxies we regard every
galaxy as a visible member of a density enhancement (group or cluster)
within the visible range of absolute magnitudes, $M_1$ and $M_2$,
corresponding to the observational window of apparent magnitudes at
the distance of the galaxy. This assumption is based on observations
of nearby galaxies, which indicate that practically all field galaxies
belong to poor groups like our own Galaxy, where one bright galaxy is
surrounded by a number of faint satellites. Using this assumption, we
find groups and clusters with their haloes, as single giant galaxies
with their companions, or groups/clusters with their faint members.
Further, we assume that the luminosity function derived for a
representative volume can be applied also for individual groups and
galaxies.

The calculation of the distances, absolute magnitudes and weight
factors of galaxies was described in detail in E03a.  When calculating
total luminosities of galaxies on the basis of their observed
luminosities we used the Schechter (\cite{S76})
function with three sets of parameters. One set is based on the SDSS
luminosity function by Blanton et al. (\cite{blanton}) and is denoted
B, the other two sets on the SDSS luminosity function found in E03a,
denoted E1 and E2. The respective values of the characteristic luminosity
$M^{\ast}$ and the shape parameter $\alpha$ are given in Table~\ref{Tab1}.

In Fig.~\ref{fig:1} we show the luminous-density weights as a function
of distance.  The weights by Blanton are rather large at large
distances; the weights by Einasto E1 and E2 are lower.  The difference
between the sets E1 and E2 is due to the fact that selection effects
influence the estimated total luminosities of clusters and
superclusters in different ways.  The weights E2 have been derived
with the aim to get the correct total luminosity of the sample as a
whole at a given distance from the observer; this set yields
luminosities of superclusters independent of distance. However, in
this case the visible clusters have to include also luminosities of
the invisible clusters, and luminosities of individual clusters are
too high at large distances.  To avoid this distortion of cluster
luminosities we have used the weights of the set E1; this set yields
for clusters the estimated total luminosities, which are statistically
independent of their distance (for details see E03a).  The right panel
of Fig.~\ref{fig:1} shows the observed and total luminosities of
galaxies at various distances.

The next step is the search for groups and clusters of galaxies.  Here
we used the conventional friends-of-friends algorithm by Zeldovich,
Einasto \& Shandarin (\cite{zes82}, hereafter ZES). Another algorithm
was suggested by Huchra \& Geller (\cite{hg82}, hereafter HG).  These
algorithms are essentially identical with one difference: ZES used a
constant search radius to find neighbours whereas HG applied a
variable search radius, depending on the volume density of galaxies at
a particular distance from the observer.  We compared SDSS cluster
catalogues obtained by both algorithms, and found that mean virial
radii of groups/clusters are practically constant for the constant
search radius, and increase with distance for the variable search
radius (for a comparison of group radii for both algorithms see
Einasto \etal\ \cite{e04a}).  In the following analysis we have used
only the group/cluster catalogue found with a constant search
radius. The number of groups/clusters found for both equatorial slices
is given in Table~\ref{Tab1}.

\subsection{N-body models}

We used a flat cosmological model with the parameters derived from a
joint analysis of the WMAP microwave background experiment and SDSS
data by Tegmark \etal\ (\cite{t04}) (see also Bennett \etal\
\cite{benn03}).  We calculated three models with cube sizes $L = 100$
and 200 \Mpc. The smaller cube was calculated for a $128^3$ mesh and
the same number of DM particles, and the two larger cubes for a
$256^3$ mesh and particles; we designate these models as M100, M200A,
and M200B, respectively.  The cosmological parameters of models are
given in Table~\ref{Tab2}; here $\Omega_m$ is the matter density (dark
plus baryonic matter), $\Omega_{\Lambda}$ is the dark energy density
(all in units of the critical cosmological density), $\sigma_8$ is the
present density fluctuation parameter, and $M_p$ is the mass of a
single particle. Here and elsewhere $h$ is the present-day
dimensionless Hubble constant in units of 100 km s$^{-1}$
Mpc$^{-1}$. For the models M100 and M200A the initial power spectrum
was taken using the approximation formula given by Klypin et
al. (\cite{klypin93}).  For the model M200B the initial power spectrum
was generated using the COSMICS code by Bertschinger ({\tt
http://arcturus.mit.edu/cosmics}); here we accepted the baryonic
matter density $\Omega_b= 0.044$.

In simulations we used the Multi Level Adaptive Particle Mesh (MLAPM)
code by Knebe \etal\ (\cite{knebe01}).  This code uses an adaptive
mesh technique in regions where the density exceeds a fixed threshold.
The DM-haloes were found using the conventional FoF algorithm with a
constant search radius for haloes of density contrasts $\delta n/n =
80$, 125 and 411; these correspond to neighbourhood radii $b=0.23$,
0.20 and 0.134 in units of the mean particle separation, respectively.
The density contrast $\delta n/n = 80$ coincides with that used by
Tucker et al. (\cite{Tucker00}) in the search of loose groups, $\delta
n/n = 411$ describes virialized haloes in our accepted ``concordance''
model (see Peacock \cite{p99}), and the intermediate value of the
neighbourhood radius 0.2 was advocated by Jenkins et
al. (\cite{jfw01}). The DM-haloes in model M200B, used subsequently in
our analysis, were found using the neighbourhood radius $b=0.23$. In
Table~\ref{Tab2}, $N_{min}$ is the minimum number of particles in the
DM-haloes, and $N_{halo}$ is the number of haloes found.

In small DM-haloes some particles have rather large velocities
relative to the rest of the halo; these particles evidently do not
belong to the virialized part of the halo.  To avoid the inclusion of
unbound objects we should apply the virial theory condition
$E_r=E_{kin}/|E_{pot}|<0.5$ (here $E_{pot}$ is the potential energy
and $E_{kin}$ the kinetic energy of a group). However, in groups with
too high kinetic energy only a small fraction of particles are
responsible for this effect; thus by excluding all these groups we
would reduce the number of groups too much. To reduce statistically
this effect we applied in model M200B a more modest criterion,
$E_r<0.8$.  The model M200A was used only to investigate the evolution
of populations of particles of various local and global density, so for
this model individual DM-haloes were not found.

\section{SDSS DR1 clusters}

\subsection{The density field of the SDSS DR1}

The SDSS DR1 equatorial slices are very thin, thus
\mbox{3-dimensional} and 2-dimensional density fields are very similar
to each other.  Taking this into account we calculated only the
2-dimensional luminosity density fields for observational samples.  As
in E03a and E03b we calculated the high-resolution luminosity density
field using Gaussian smoothing with a rms scale of 0.8~\Mpc, and the
low-resolution field with a rms scale of 10~\Mpc.  The high-resolution
field was found using the Schechter parameters of the set E1, the
low-resolution field with the parameter set E2.

The low-resolution field yields information on large over-density
regions.  This field was used to define superclusters as connected
over-density regions. As in E03a we used density thresholds of 
1.8--2.1 to find superclusters.  At lower thresholds superclusters start
to merge into percolating systems, and thus the definition of
superclusters as the largest but still isolated high-density regions
is violated.  For higher thresholds the number of superclusters
decreases rapidly, since many of them have lower peak density.  The
catalogue of superclusters found using the SDSS DR1 with the Schechter
parameters of the set E2 is rather close to the catalogue in E03a that
used the SDSS EDR, except that the total luminosities of superclusters
vary a bit due to the use of more complete data.

\begin{figure}[ht]
\centering
\resizebox{0.45\textwidth}{!}{\includegraphics*{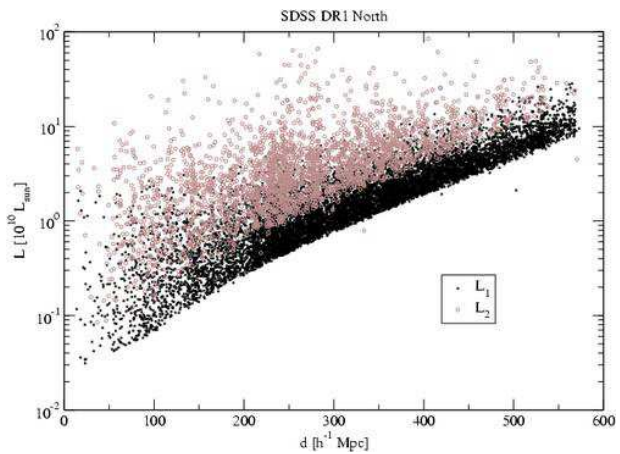}}
\hspace{2mm}
\caption{The luminosities of SDSS Northern groups/clusters at
  different distances, 
  corrected for galaxies outside the visibility window. Grey symbols
  stand for groups with at least two visible galaxies, black symbols -- for
  groups containing only one galaxy in the visibility window. The
  luminosity dependence of galaxies of the SDSS Southern slice is very
  close to that of the Northern slice.
}
\label{fig:2}
\end{figure}

\begin{figure}[ht]
\centering
\resizebox{0.45\textwidth}{!}{\includegraphics*{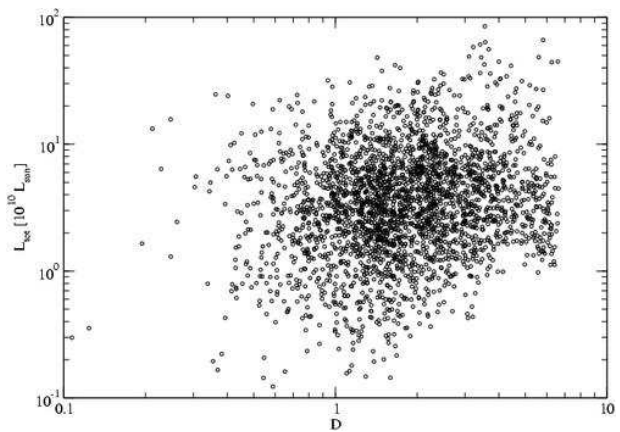}}
\hspace{2mm}
\caption{The luminosities of the SDSS DR1 Northern slice clusters as a
  function of the environmental density, found by Gaussian smoothing
  of the luminous density field with the rms scale 10~\Mpc. Data for
  the Southern slice are very similar.  }
\label{fig:3}
\end{figure}

\subsection{Properties of the SDSS clusters in various environments}

Fig.~\ref{fig:2} shows the luminosities of groups/clusters at
different distances from the observer.  We see that there exists a
well-defined lower limit of cluster luminosities at larger
distances. The limit is linear in the $\log L - d$ plot.  Such
behavior is expected as at large distances an increasing fraction of
clusters do not contain any galaxies bright enough to fall into the
observational window of absolute magnitudes, $M_1 \dots M_2$.  The
limit is lower for groups containing only one galaxy in the visibility
window; these groups are systems like our Local group with one bright
galaxy surrounded by faint companions. The difference in the
low-luminosity limits of groups containing at least one or two
galaxies in the visibility window is by a factor of two as expected
(this factor corresponds to the case when two galaxies in the
visibility window have the same luminosity).

\begin{figure*}[ht]
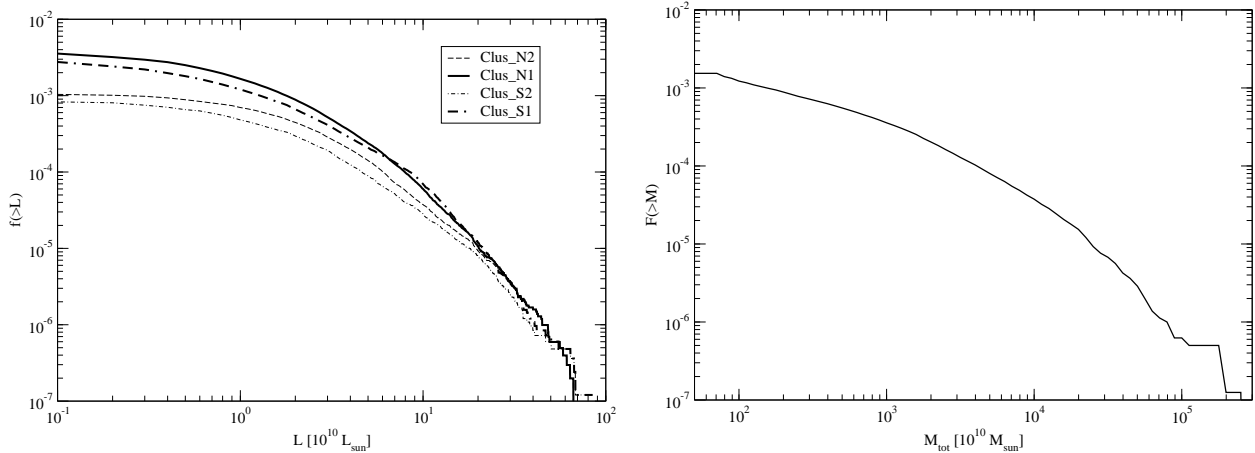

\centering
\resizebox{0.45\textwidth}{!}{\includegraphics*{einasto_fig4a.eps}}
\hspace{2mm}
\resizebox{0.45\textwidth}{!}{\includegraphics*{einasto_fig4b.eps}}
\hspace{2mm}
\caption{Left: the luminosity functions of the SDSS DR1
  groups/clusters. Thin lines show the luminosity functions found
  using the clusters with at least two galaxies in the observational
  window, bold lines show luminosity functions for all
  groups/clusters, including groups with only one galaxy in the
  observational window. When calculating the total luminosities of
  groups, the Schechter function parameters of the set E1 were
  used. Right: the cumulative function of the total masses of
  DM-haloes of the model M200B.  }
\label{fig:4}
\end{figure*}

Let us compare now properties of groups/clusters in various
environments. We shall use the density found with the 10~\Mpc\
smoothing as the global density in the supercluster environment of
clusters.  The environmental effect is shown in Fig.~\ref{fig:3}.
There is a correlation between the luminosity of the most luminous
clusters and their environmental density.  The most luminous clusters
in high-density regions have a luminosity a factor of about 5 higher
than the most luminous clusters in low-density regions.  Using a
different definition of the large-scale environment, a similar effect
was found in the vicinity rich clusters of galaxies by E03c and E03d,
and in the vicinity of massive DM-halos by E04b.

\subsection{Cluster luminosity functions}

Fig.~\ref{fig:4} shows the integrated luminosity functions of the
groups/clusters of the SDSS DR1 Northern and Southern samples.  The
absence of low-luminosity clusters at large distances has been taken
into account by the standard $V_{max}^{-1}$ weighting procedure (for
details see E03a).  The luminosity function was calculated separately
for groups/clusters with at least two visible galaxies, and for all
groups/clusters including the systems with only one visible galaxy in
the visibility window. In both cases the numbers of clusters have been
corrected for selection effects.  We see that in the second case the
number of groups/clusters per unit volume is larger by a factor of
$\sim$3 for the low luminosity section of the luminosity function.
This result shows that the groups with one bright main galaxy dominate
among low-luminosity groups.  We believe that this higher density
represents the true number-density of low-luminosity groups better
than the density found from groups with at least two galaxies in the
observational window.  It is well known that in our vicinity the
majority of groups are similar to our Local Group, which consists of
two subgroups with one bright main galaxy (our Galaxy and M31) and a
number of considerably fainter companion galaxies.

\begin{figure}[ht]
\centering
\resizebox{0.45\textwidth}{!}{\includegraphics*{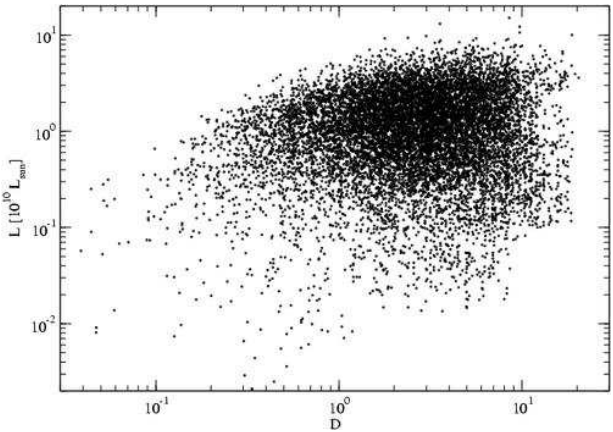}}
\hspace{2mm}
\caption{The luminosities of galaxies in various density environments for
  the SDSS EDR Southern slice. The environmental density was calculated using
  Gaussian smoothing with the rms scale of 2~\Mpc.
}
\label{fig:5}
\end{figure}

\begin{figure*}[ht]
\centering
\resizebox{0.45\textwidth}{!}{\includegraphics*{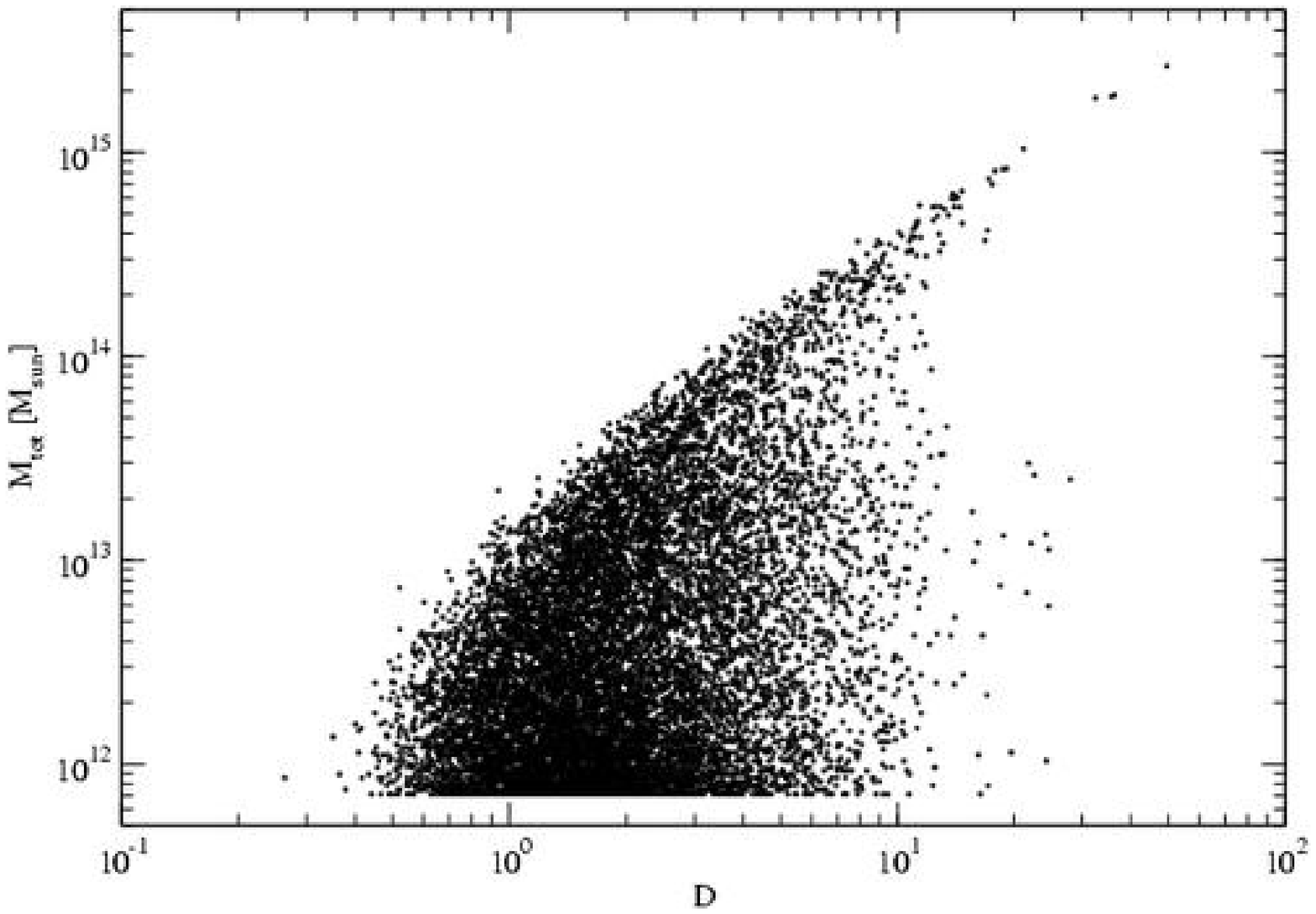}}
\hspace{2mm} 
\resizebox{0.45\textwidth}{!}{\includegraphics*{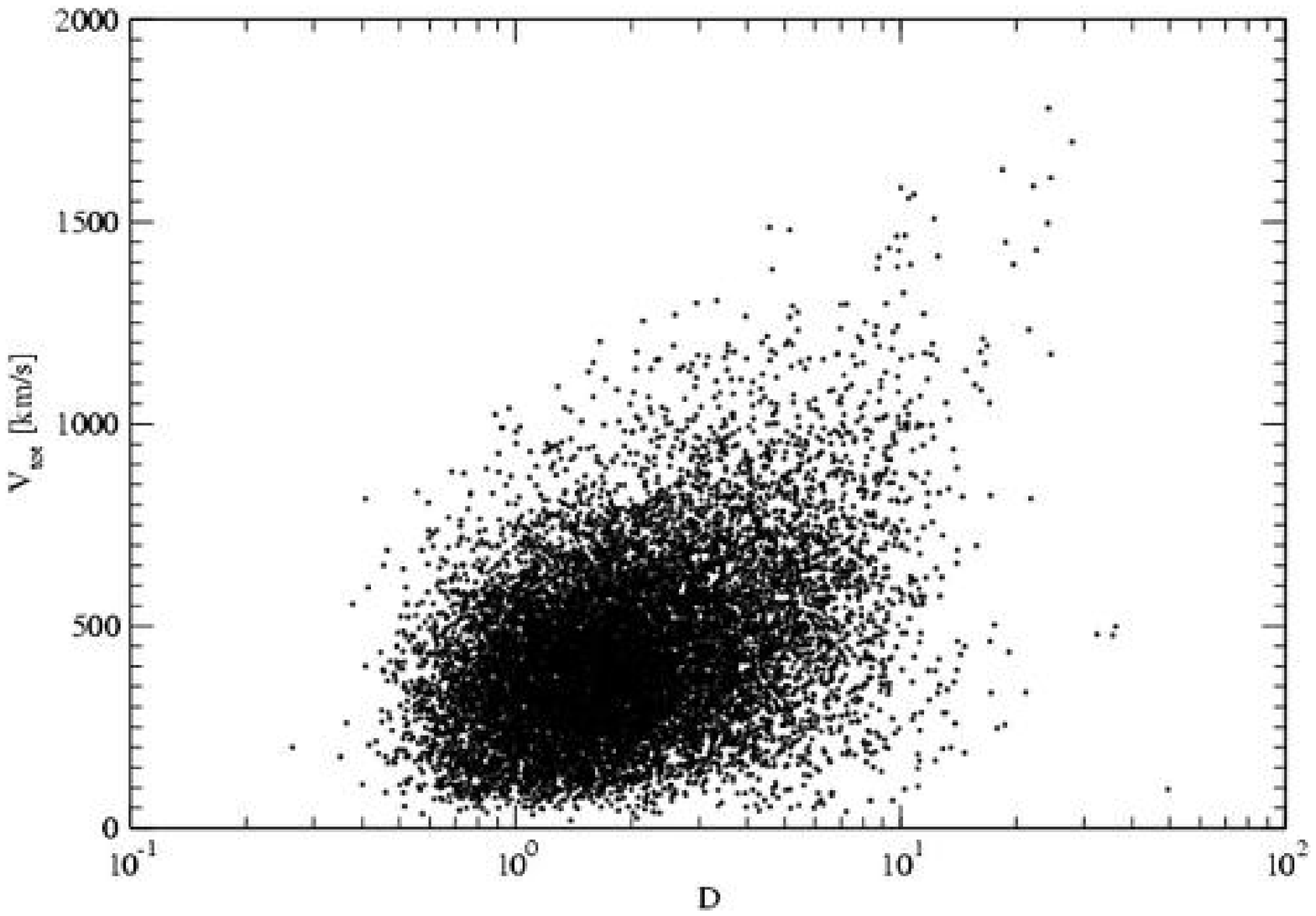}}
\hspace{2mm}
\caption{The dependence of the parameters of DM-haloes in the model
  M200B on the density of the environment, $D$ (in units of the mean
  density). The density of the environment was determined from the
  low-resolution field using the Epanechnikov kernel with a radius of
  8~\Mpc. Left panel: total mass; right panel: full velocity.  }
\label{fig:6}
\end{figure*}

\subsection{Luminosities of galaxies in different environments}

It is well known that in clusters brighter galaxies are concentrated
toward cluster centres.  For this reason the smoothing scale must be
relatively small in order to be sensitive to the positions of galaxies
within clusters. Taking this into account we shall use the density
found with the 2~\Mpc\ smoothing as an environmental parameter to
describe the surrounding density of galaxies.  The luminosity of
galaxies as a function of the environmental density is shown in
Fig.~\ref{fig:5}; here we used the SDSS EDR to calculate the
luminosity density.  This figure shows that the most luminous galaxies
in high-density regions are about 5--10 times more luminous than the
most luminous galaxies in low-density environments.  It is interesting
that the contrast in luminosity between the high- and low-density
regions is of the same order than the luminosity contrast of
groups/clusters between the high- and low-density regions.

\section{Clusters in simulations}

\subsection{DM haloes and density fields}

The second goal of our study is comparison of observational data with
numerical simulations.  We have used for this purpose three
simulations, M100, M200A and M200B, with $128^3$ and $256^3$ particles
in 100~\Mpc\ and 200~\Mpc\ cubes, respectively.  The cosmological
parameters of the models are given in Table~\ref{Tab2}.  In the
analysis that follows we used the DM-haloes identified by the FoF
algorithm with the search radius 0.23 (in units of the mean particle
separation) for the model M200B.

The density fields for the N-body simulations were found using all
particles of simulations, i.e. we calculated the simulated true total
matter density fields. The high-resolution density was found using the
conventional cloud-in-cell (CIC) scheme and additional Gaussian
smoothing with the rms scale 0.8 in grid cell units (0.6~\Mpc).  
Gaussian smoothing was used in order to avoid the presence of
numerous empty cells in low-density regions.  The low-resolution
density field was calculated using an Epanechnikov kernel with the
radius 10 in grid units, which corresponds to 8~\Mpc.  This field was
applied to find simulated superclusters and environmental densities of
DM-haloes.

{\scriptsize
\begin{table*}
      \caption[]{The threshold density, volume and mean density  of various
       environments}

         \label{Tab3}
      \[
         \begin{tabular}{c|ccc|cccc|cccc}
            \hline
            \noalign{\smallskip}
            Sample &\multicolumn{3}{c|}{Threshold}&& Volume &&&& Density && \\
 & $D_0$ &$D_1$ & $D_2$  & Void &P fil & R fil & Scl & 
             Void &P fil & R fil & Scl \\  

            \noalign{\smallskip}
            \hline
            \noalign{\smallskip}
$z=5$ & 0.706&0.845&1.035&17.3&20.8&27.1&34.8&0.548&0.732&0.929&1.442 \\
$z=2$ &0.587&0.828&1.239&25.8&24.8&26.3&23.0&0.374&0.619&0.955&2.166 \\
$z=1$ &0.527&0.857&1.552&32.8&26.9&24.4&15.9&0.289&0.566&1.035&3.151 \\
$z=0$ &0.484&1.000&2.661&46.7&26.7&19.1& 7.5&0.207&0.549&1.350&6.641 \\
            \noalign{\smallskip}
            \hline
         \end{tabular}
      \]
   \end{table*}
}

\subsection{Properties of DM-haloes in different environments}

The dependence of the total mass of DM-haloes on the density of the
environment is shown in the left panel of Fig.~\ref{fig:6}.  Here the
dependence of the DM-halo mass on the density of the environment is
very well expressed: the most massive clusters in a high-density
environment are by a factor of a hundred more massive than the most
massive clusters in a low-density environment.  For a numerical
simulation, we have also information on the velocities of DM-haloes,
which is absent for the SDSS DR1 groups/clusters.  The right panel of
Fig.~ \ref{fig:6} shows the dependence of the full cluster velocity on
the density of the environment.  The full velocity has a less
pronounced density dependence, but here, too, DM-haloes in most dense
environments have a factor of ten larger velocities than DM-haloes in
less dense environments.

A similar result was obtained by E04b using a different definition of
the density of the environment (the distance to the 5th nearest
neighbour).  In this paper, in addition to the correlations considered
here, the virial radii and the intrinsic rms velocities of DM-haloes
in various environments were also studied.  The virial radii of
DM-haloes in a high-density environment are larger than those in a in
low-density environment, but here the contrast between the high- and
low-density regions is not so large. The rms velocities of DM-haloes
have a very strong environmental dependence, similar to the dependence
observed for masses. This is natural, as the rms velocities of
virialized haloes are determined by their masses.

\section{The evolution of various environmental regions}

To understand better the dependence of cluster properties on the
environment we determined the distribution of particle densities in
regions of various environmental density.  For this purpose, for the
model M200A we found for every particle in the simulation two density
values, the local and the global density. The local density attributed
to the particle was found as described above with Gaussian smoothing
with the rms scale 0.8 in grid cell units (0.6~\Mpc).  The global density
was found using the low-resolution density field as described above
(smoothed with the Epanechnikov kernel of the radius 8~\Mpc). The
density fields and particle densities were found for four epochs of
the simulation, corresponding to the redshifts $z=5, ~2,~1, ~0$.  The
simulations started at the redshift $z=50$, so at all epochs
considered in our analysis the density field was well evolved.

We divided the whole simulation volume into four regions by increasing
global density. These regions correspond to voids, poor and rich
filaments, and superclusters.  The analysis of the distribution of
particles for the present epoch $z=0$ shows that approximately 50\%~
of all particles are presently located in rich supercluster regions
with the global density $D>2.661$. Rich filaments (actually poor
superclusters) can be localised as systems lying in the range of
global densities $1 < D \leq 2.661$; at the present epoch about 25\%
of particles lie in this density range.  Poor filaments can be found
as systems in the global density range $0.484 < D \leq 1$; about 15~\%
of particles lie presently in this density range. The rest of
particles at global densities $D \leq 0.484$ constitute the void
region; the fraction of particles in this density range is at the
present epoch about 10\%.

If we are interested in the dynamics of different individual regions
we have to trace back the positions of individual particles.  This
approach has been followed by Gottl\"ober et al. (\cite{gottl03}) in
their study of the evolution of individual voids. Our task is simpler,
as we are interested in the evolution of the simulation sample as a
whole. It is clear that during the evolution DM-particles cluster
locally to form DM-haloes, but most particles do not change their
large-scale environment.  The larger the scale, the smaller are the
velocities at that scale.  In other words, we may assume that the same
fraction of particles presently located in the region of the highest
global density was in earlier epochs also in the regions of the
highest global density, and similarly the fractions of particles in
other ranges of global density did not change. Under this assumption
we found for each simulation epoch threshold global densities $D_i$
which divide the sample of all particles at a given epoch into regions
of global density which occupy, from lowest values upward, 10\%, 15\%,
25\%, and 50\% of all particles. As noted above, we call the
respective regions the void regions, the poor and rich filament
regions, and the supercluster regions. The respective global density
thresholds for all epochs considered are given in Table~\ref{Tab3}.

\begin{figure*}[ht]
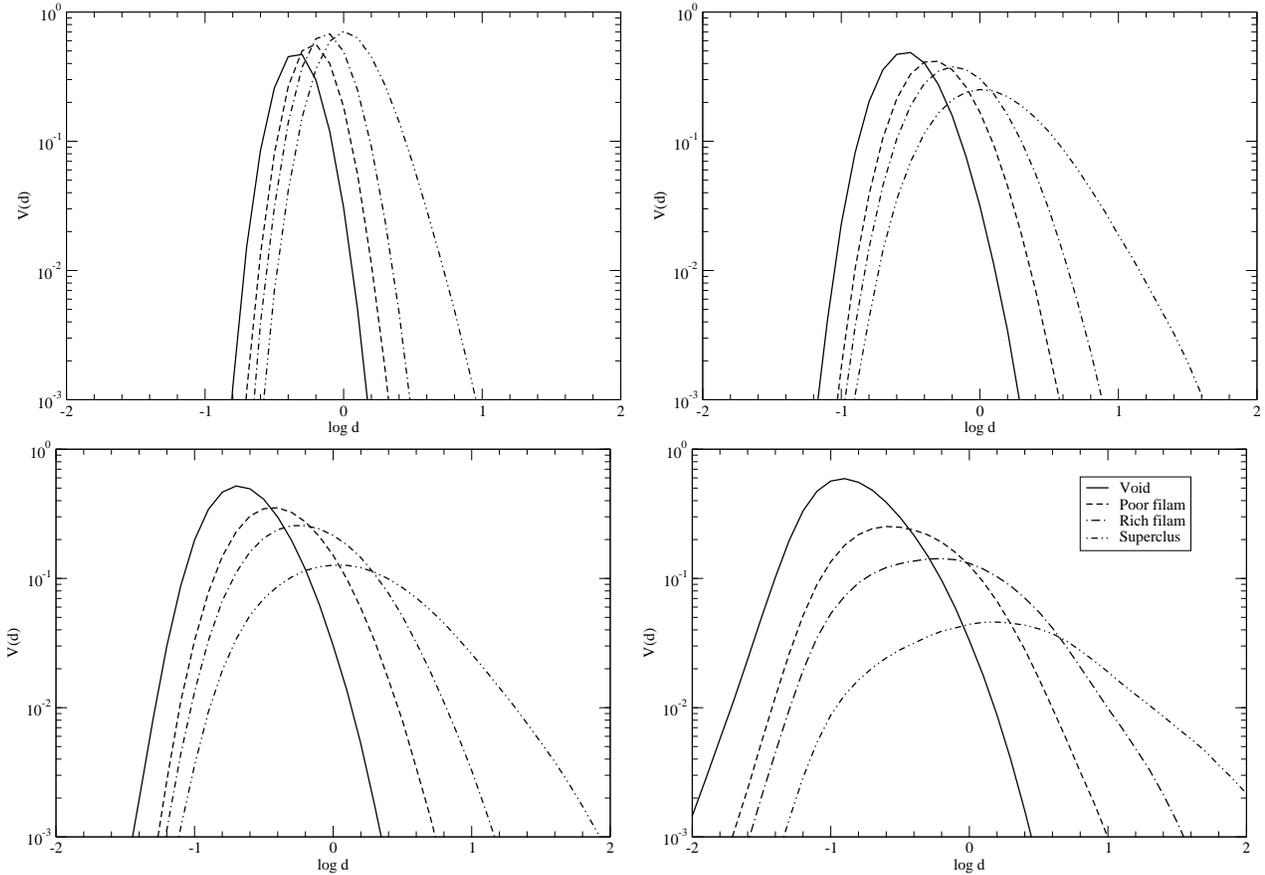

\centering
\resizebox{0.45\textwidth}{!}{\includegraphics*{einasto_fig7a.eps}}
\hspace{2mm} 
\resizebox{0.45\textwidth}{!}{\includegraphics*{einasto_fig7b.eps}}
\hspace{2mm}
\resizebox{0.45\textwidth}{!}{\includegraphics*{einasto_fig7c.eps}}
\hspace{2mm} 
\resizebox{0.45\textwidth}{!}{\includegraphics*{einasto_fig7d.eps}}
\hspace{2mm}
\caption{The distribution of volume fraction as a function of local density
  $d$ in regions of various environment in model M200A. The
  distribution is found for various regions of global density which
  correspond to voids, poor and rich filaments, and superclusters,
  containing 10\%, 15\%, 25\%, and 50\% of all particles in regions of
  increasing global density $D$, respectively.  The upper panels are
  for the epochs $z=5$, and $z=2$, and the lower panels for the epochs $z=1$
  and $z=0$ (from left to right).
}
\label{fig:7}
\end{figure*}

During their dynamical evolution, superclusters shrink in volume and
voids expand. To follow the change of the volume of various
environmental regions we give in Table~\ref{Tab3} the volume occupied
by various regions (in per cents of the total simulation volume). The
distribution of volume fraction as a function of the local density $d$
is shown in Fig.~\ref{fig:7}, for all epochs of simulation.  Different
lines show the distribution for various global density $D$ ranges: the
void, the poor and rich filament, and the supercluster range.  The
fraction of particles in each region is known, so we can calculate the
mean density of matter in each region (in units of the overall mean
density); these mean densities are also given in the table. We see
that void regions occupy initially (at $z=5$) only 17\% of the volume,
during evolution this fraction grows to 47\%, and the mean density
shrinks from 0.5 to 0.2. Fig.~\ref{fig:7} shows that in lowest density
regions the density is at present epoch less than 0.01. On the other
hand, the 50\% of all matter that lies in high-density regions
(superclusters) occupies initially 35\% of space, this fraction
decreases to 8\%, and respectively the mean density increases from 1.4
to 6.6; in the highest density regions it is higher than 100.  The
poor and rich filament regions have intermediate behaviour, their
volume fractions and mean densities changing much less.

\begin{figure*}[ht]
\centering
\resizebox{0.45\textwidth}{!}{\includegraphics*{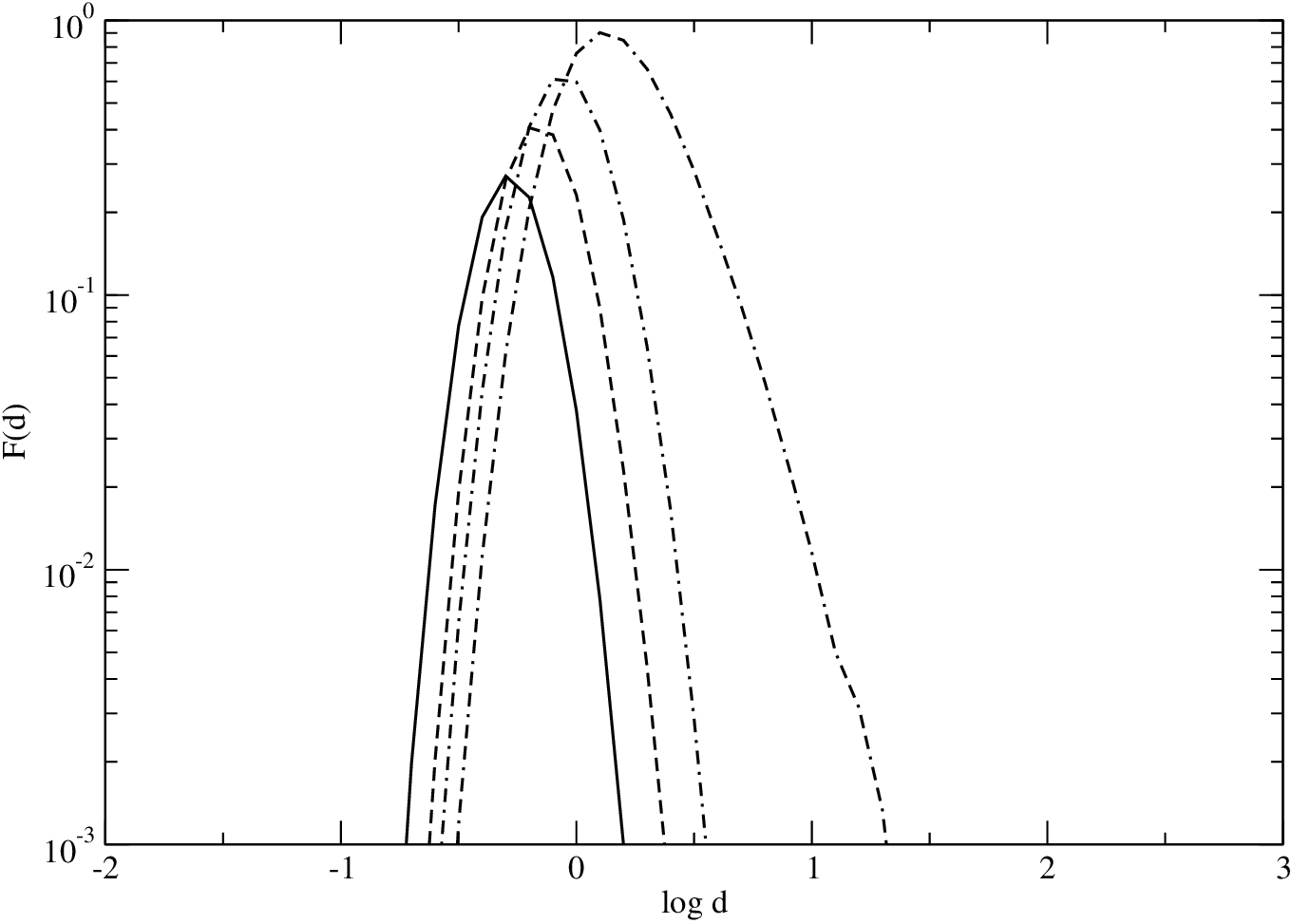}}
\hspace{2mm} 
\resizebox{0.45\textwidth}{!}{\includegraphics*{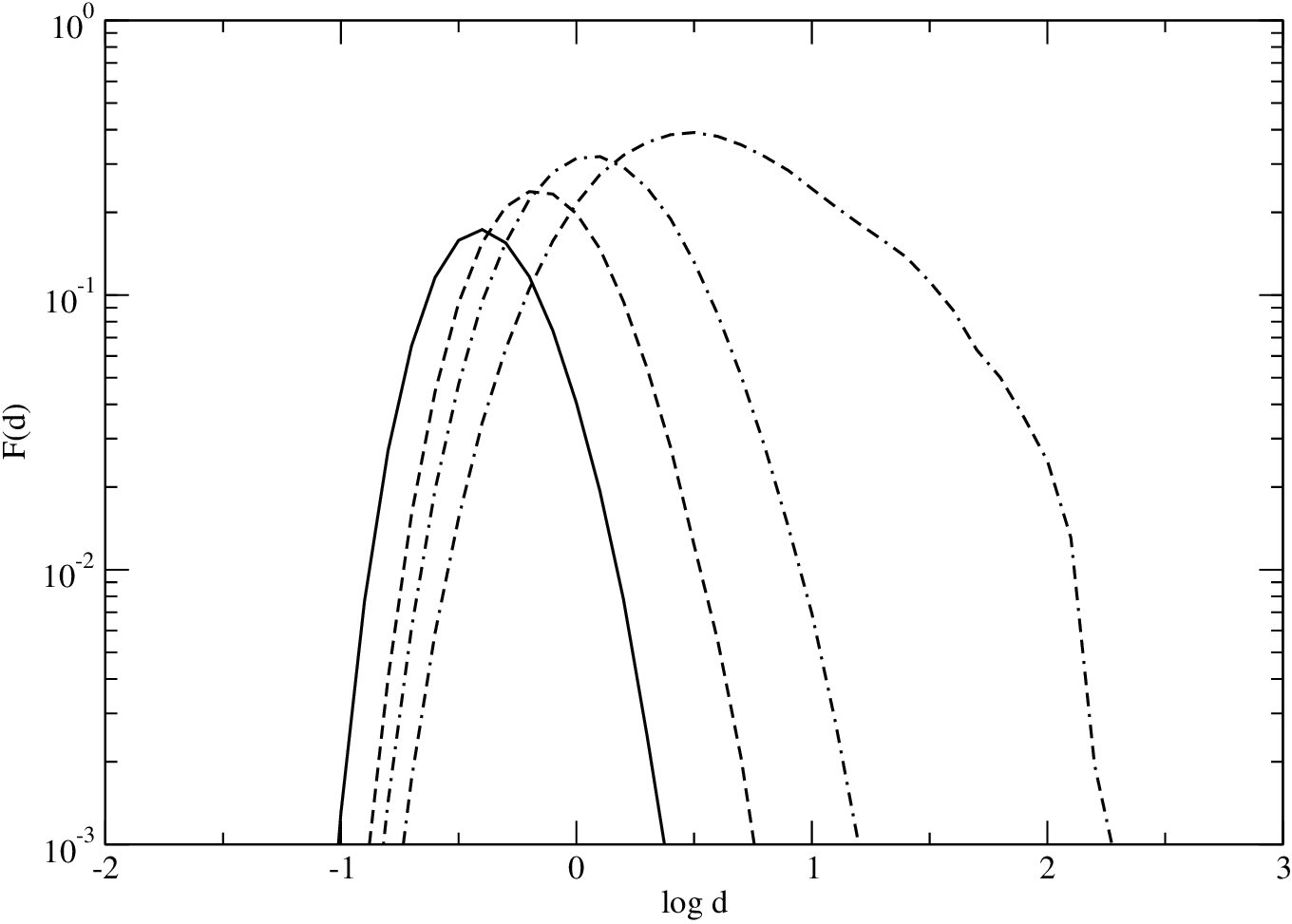}}
\hspace{2mm}
\resizebox{0.45\textwidth}{!}{\includegraphics*{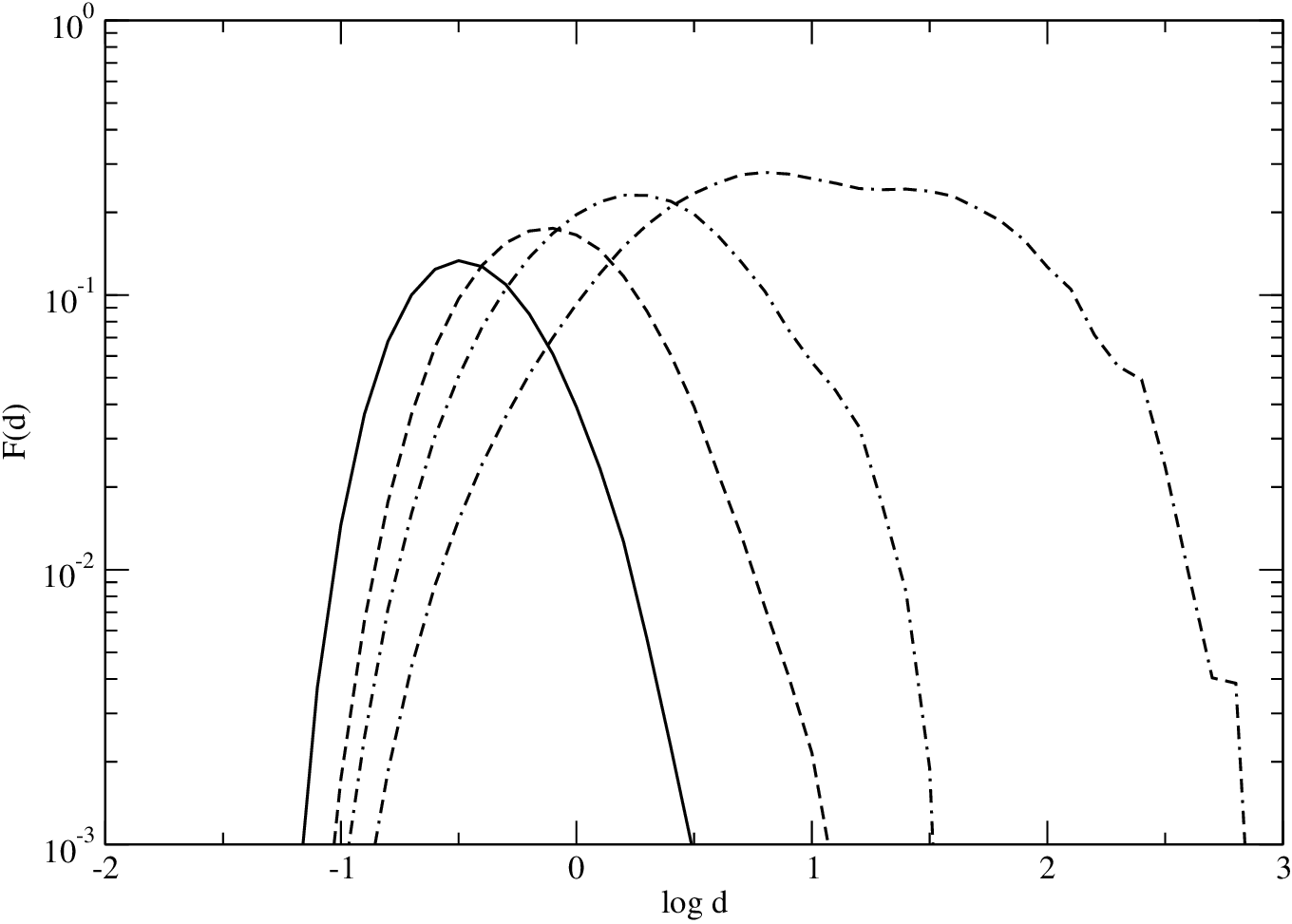}}
\hspace{2mm} 
\resizebox{0.45\textwidth}{!}{\includegraphics*{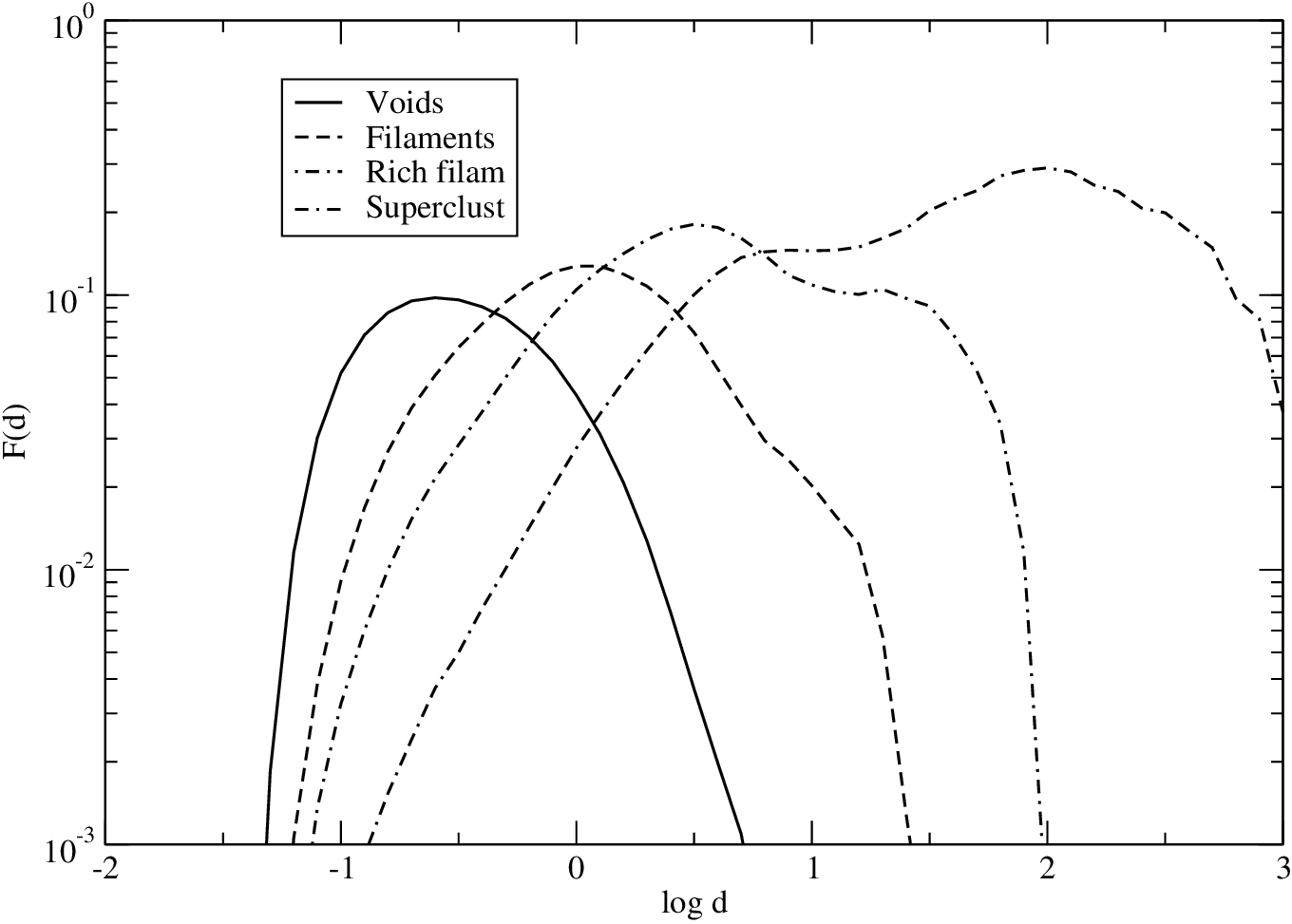}}
\hspace{2mm}
\caption{The distribution of particles as a function of local density
  $d$ in regions of various environment in the model M200A. The
  distribution has been found for various regions of global density which
  correspond to voids, poor and rich filaments, and superclusters,
  containing 10\%, 15\%, 25\%, and 50\% of all particles in regions of
  increasing global density $D$, respectively.  The upper panels show
  the epochs $z=5$, and $z=2$, and the lower panels show the epochs $z=1$
  and $z=0$ (from left to right).
}
\label{fig:8}
\end{figure*}

{\scriptsize
\begin{table*}
      \caption[]{The fraction of particles of various
      local density in different environments}

         \label{Tab4}
      \[
         \begin{tabular}{c|ccc|ccc|ccc|ccc}
            \hline
            \noalign{\smallskip}
            Sample &\multicolumn{3}{c|}{Voids}&
            \multicolumn{3}{c|}{Poor filaments}&
	    \multicolumn{3}{c|}{Rich filaments}&
            \multicolumn{3}{c}{Superclusters} \\  
	    & 0 & 1 & 2 & 0 & 1 & 2 & 0 & 1 & 2 & 0 & 1 & 2 \\
            \noalign{\smallskip}
            \hline
            \noalign{\smallskip}
$z=5$ &8.4&1.1&0.0&10.4&4.8&0.0&11.6&13.5&0.0&8.9&40.1&1.1\\ 
$z=2$ &8.3&1.4&0.0& 8.8&6.5&0.1&8.1&16.1&1.0&4.8&26.4&18.7\\
$z=1$ &7.9&1.6&0.0& 7.4&7.4&0.4&5.8&15.3&4.2&2.6&16.8&30.7\\
$z=0$ &7.4&2.2&0.0& 5.0&8.1&1.5&3.0&12.9&9.9&0.8& 8.8&40.3\\
            \noalign{\smallskip}
            \hline
         \end{tabular}
      \]
   \end{table*}
}

We determined in each of the regions the distribution of the particles
by their local density $d$.  Using the local density, we can assign
particles to different populations.  We emphasize that particles with
local densities less than 1 cannot belong to clusters or groups, since
galaxy formation starts only when the local density exceeds a certain
critical threshold much higher than the mean density (see Press \&
Schechter \cite{ps74}).  Following these ideas we classify the
population of particles with local density below unity as primordial
(non-clustered) particles, the population of particles with the local
density values in the range $1 \leq d <10$ as poor cluster (group)
particles, and the population of particles with the local density
values $d > 10$ as rich cluster particles.  For an illustration of the
distribution of particles of populations with various local densities
see Fig.~1 of Einasto et al.  (\cite{1999ApJ...519..456E}).

The distribution of the number of particles according to their local
density $d$ is given in Fig.~\ref{fig:8} for all simulation epochs
considered. The Figure is similar to the previous one, only here we
plot the distribution of mass instead of volume.  The Table~\ref{Tab4}
gives the fractions of particles (in per cents of the total number of
particles in simulation), which belong to populations of various local
density: the populations of the primordial, the poor cluster and the
rich cluster particles; these populations are denoted as 0, 1 and 2,
respectively.  The fractions are given, as explained above, separately
for the void, the poor filament, the rich filament, and the
supercluster environment.
 
Let us discuss the data in more detail. Consider, first, the void
regions.  The void particle population distributions are drawn in
Figs.~\ref{fig:7} and ~\ref{fig:8} by solid lines (see also the
columns labeled 0 in Table~\ref{Tab4}).  We see that at all epochs the
distribution of local densities of void particles is rather
symmetrical, both in volume and mass.  In void regions most particles
belong at all epochs to the primordial non-clustered population 0. At
the epoch $z=5$ about 1\% of the particles belong to the poor cluster
population (the columns labeled 1 in Table~\ref{Tab4}); this fraction
increases with time and reaches 2.2\% at the present epoch ($z=0$).
Initially there are no particles of rich cluster population among void
particles (columns 2), and at the present epoch a very tiny fraction
of particles has crossed the threshold of the rich cluster population.

The distribution of particles in the poor filament regions is plotted in
Figs.~\ref{fig:7} and ~\ref{fig:8} by dashed lines. We see that the
initial distribution of particles is similar to the distribution in
void regions; however, the distributions are shifted toward higher
local density values, and thus the fraction of the poor cluster
population 1 is higher. The poor cluster population 1 grows with time,
so that at the present epoch about half of the poor filament particles 
belong to it.  The rich cluster population 2 fraction in these regions
grows from zero to 1.5\%.  Most of the volume in the poor filament
region is occupied by local voids with local density values $d< 1$
(see Fig.~\ref{fig:7}).

The evolution of particles in the rich filament environment is shown
in Figs.~\ref{fig:7} and ~\ref{fig:8} by dot-dashed lines.  In this
region initially about half of the particles belong to the primordial
population 0, and the other half to the poor cluster population 1; there
are no particles of the rich cluster population 2. As time goes by,
the fraction of primordial particles rapidly decreases and the
fraction of rich cluster particles increases; the fraction of poor
cluster particles changes little.

For the evolution of the supercluster region see the dot-dash-dashed lines
in Figs.~\ref{fig:7} and ~\ref{fig:8}. In this environment the
fraction of primordial particles 0 rapidly decreases with time to almost
zero, the fraction of poor cluster particles 1 decreases from about 40\%
to 9\%, and the fraction of rich cluster population 2 increases from 1\%
to 40\% (from $z=5$ to $z=0$).  A large fraction of particles have
local densities far in excess of our limiting density $d=10$. However,
a considerable fraction of space is still occupied by local voids, as
seen from Fig.~\ref{fig:7}.

These results can be summarized as follows: in void regions the mean
density decreases continuously, as a result DM-haloes almost do not
evolve dynamically, and most particles remain as primordial
(non-clustered) ones.  In supercluster regions the dynamical evolution
is very rapid, the primordial population clusters rapidly, and later
evolution consists of the transition of galaxies and groups to rich
clusters. Here we have not followed the evolution of individual clusters,
but it is clear that in this later stage merging of smaller DM-haloes
to form rich DM-haloes plays an important role. In poor and rich
filament regions the evolution is between these two extreme cases.

\section{Discussion}

\subsection{Luminosity/mass functions in real and simulated clusters}

Let us compare first the luminosity and mass functions of real and
simulated groups/clusters.  In our preliminary study (E03a and E03b)
we defined groups and clusters as enhancements of the 2-dimensional
high-resolution luminosity density field.  In the present work we used full
3-dimensional data to define groups/clusters, both for the real data and
simulations.

New more accurate data show (see Fig.~\ref{fig:4}), that there is no
essential difference between the luminosity functions in the Northern
and Southern strip of the survey.  Our preliminary analysis based on
the SDSS EDR and groups/clusters defined using the 2-dimensional
luminosity density data suggested the presence of differences between
the Northern and Southern strips.  The new analysis does not support
our previous conclusion.

The volume density of groups/clusters according to the SDSS DR1 data
is $3 \times 10^{-3}$ (\Mpc)$^{-3}$ for $L \geq 10^9$ $L_{\odot}$
groups/clusters.  This estimate is in fairly good agreement with the
estimates of the number density of groups based on the group mass
function by Girardi \& Giuricin (\cite{gg00}), see also Hein\"am\"aki
et al. (\cite{hei03}).

The luminosity of simulated DM-haloes is not well-defined. Thus we use
for comparison the integrated mass function of DM-haloes in
simulations, presented in the right panel of Fig.~\ref{fig:4}. We see
that the overall shape of the integrated mass function is rather
similar to the luminosity function of the real groups/clusters. In the
low-mass range the mass function of DM-haloes is steeper than the
luminosity function of the real groups/clusters. The presently available data
are insufficient to judge if this difference is significant or not; a more
detailed study of simulated samples is needed, with simulated 
galaxies generated in DM-haloes.   The mean volume density
of DM-haloes is very close to the mean volume density of groups in the SDSS
DR1, thus we can say that the population of DM-haloes represents the
population of real groups/clusters rather well (the overall bias is small).

\subsection{Environmental effects in real and simulated clusters}

Let us compare now the environmental dependence in real and
simulated cluster samples.  The environmental dependence has been
investigated using two completely independent parameters to
characterize the large-scale environment.  E03c compared groups and
clusters in high density environments, defined as a neighbourhood of
rich clusters, and in low density environments, far from rich
clusters.  E03d compared the properties of groups and clusters that
belong to superclusters and the properties of groups/clusters that do
not belong to superclusters.  E04b used the distance to the 5th nearest
neighbour as a parameter of the large-scale environment to describe
the environment of DM-haloes in simulations.  E03a and E03b used for
this purpose the low-resolution luminosity density field, as we do in
the present study.  

These parameters of the large-scale environment used in various
studies are independent of each other and characterize the environment
from different points of view.  The luminosity density field approach
takes into 
account the luminosity or the mass of neighbouring objects, including
the dark matter particles in simulations.  The nearest neighbour approach,
as well as the proximity to rich clusters, depends only on the number
and position of neighbours. Their luminosity or mass is ignored.

All relations considered so far between the physical parameters of
groups/clusters and the environmental parameter show the presence of
well-defined correlations, both in real as well as in simulated
samples: in high-density regions (superclusters) groups/clusters are
brighter and more massive than in low-density regions (void regions).
Also they have slightly larger radii and greater bulk velocities (see
E04b).

\subsection{Comparison to previous work}

Historically, the dependence of galaxy properties on their large-scale
environment has been investigated long ago, starting from the pioneering
studies by Davis \& Geller (\cite{dg76}) and Dressler (\cite{dr80}).
In these early studies a striking contrast between 
the morphological types of galaxies in the cluster and field
environments was found: in 
clusters elliptical galaxies dominate and in the field dominate spiral and
irregular ones. More recently Einasto (\cite{e91a},~\cite{e91b}) and
Mo et al. (\cite{mexd92}) found the dependence of galaxy luminosity
and morphological type on the large-scale environment up to the scale
of 10~\Mpc. This result was confirmed by Lindner et al. (\cite{l95},
~\cite{l96}) by the study of voids defined by galaxies of different
absolute magnitude: bright galaxies define much larger voids than
intrinsically faint galaxies; these faint galaxies form poor filaments
inside large voids surrounded by bright galaxies.

Recently Balogh et al. (\cite{balogh04}), and Blanton et
al. (\cite{blant04a}, \cite{blant04b}, see also references in these
papers) investigated the dependence of physical properties of galaxies
in different local and global environment. Among physical properties
they considered colours, luminosities, H$\alpha$ emission, and the
S\'ersic (\cite{sersic68}) luminosity radial profile index.  The local
environment was defined as the spatial density on the 0.5--1~\Mpc\
scale, the global density on the 5--10~\Mpc\ scale. Blanton et
al. (\cite{blant04b}) come to the conclusion that the blue galaxy
fraction, and the recent star formation history in general, depend
mostly on the local environment.  On the other hand, Croton et
al. (\cite{cr04}) find that the luminosity function of galaxies depends
strongly on the global environment (see below).

Peebles (\cite{p01}) compared void and cluster galaxies and confirmed
the presence of a striking difference of the properties of these
galaxies.  He argued that this difference may be a challenge to the
$\Lambda$CDM model of structure formation.  Mo et al. (\cite{mo04}) came
to the conclusion that the differences in luminosity observed in regions
of different environmental density can be attributed to the differences in
the mass of DM-haloes from which galaxies form.  As our present study
shows, the masses themselves depend on the environment.

The luminosity dependence of galaxy clustering was investigated by
Norberg et al. (\cite{n01}, \cite{n02}), using the 2dF Galaxy
Redshift Survey.  They found that the clustering amplitude increases with
absolute magnitude, confirming earlier results by Einasto, Klypin \& Saar 
 (\cite{eks86}), Bromley et al. (\cite{bplk98}), and Beisbart \&
Kerscher (\cite{bk00}).  

The luminosity function of galaxies in different environment was
recently investigated by H\"utsi et al. (\cite{h02}), De Propris et
al. (\cite{dp03}) and Mo et al. (\cite{mo04}). H\"utsi et
al. estimated the luminosity function in three regions of global density,
defined by the 10~\Mpc\ smoothing, and found that the characteristic
luminosity of galaxies increases by a factor of 1.5 from the
low-density to the high-density regions.  Mo et al. argued that this
difference in the characteristic luminosity is in agreement with their
model prediction, which assumes that the Schechter approximation is
valid in virtually all environments.  

A very detailed study of luminosity functions of galaxies in
regions of different density of the large-scale environment was made
by Croton et al. (\cite{cr04}), using the full dataset of the 2dF
Galaxy Redshift Survey. Using densities smoothed on a scale of 8~\Mpc\, as
in the present paper, he divided the volume under study into 7 regions
of various density of the environment, from extreme void to cluster
populations. In all regions the luminosity function was calculated and
the parameters of the Schechter function were found.  The bright end of
the function depends primarily on the characteristic absolute
magnitude $M^*$ of the Schechter function. For the extreme void population
this parameter is $-18.3$, and for the cluster population
$-20.1$. In other words, the brightest galaxies in voids are approximately
5 times fainter than in clusters.  This result is in very good
agreement with our data on the distribution of galaxy luminosities of
the SDSS in various environments (see Fig.~\ref{fig:5}).

The dependence of the total luminosities and masses of galaxy systems
on the density of the environment has been investigated only recently
by E03a, E03b, E03c, E03d, E04a and E04b (for a discussion about the
properties of haloes in different environments see references in this
paper). These studies show a tendency similar to that of galaxy
luminosities -- in high-density regions massive and luminous clusters
dominate, whereas in low-density regions all galaxy systems are poor
and faint. The most luminous clusters in a high-density environment
are a factor of 5--10 more luminous than the most luminous clusters in
a low-density environment.  It is striking that this factor is in the
same range as for the most luminous galaxies in high- and low-density
environments.

The properties of groups of galaxies in the vicinity of rich clusters were
compared with the properties of ordinary groups by Ragone et
al. (\cite{ragone04}). They used a sample of groups identified in the
2dF Galaxy Redshift Survey, and compared the properties of groups in the
vicinity of rich clusters and the rest of the sample.  The observational
results were compared with simulated clusters using the Virgo
Consortium Simulation. In both the real and simulated groups there exist
similar relations: the larger the host mass, the higher is the 
luminosity or the mass of the DM-halo.

Comparison of the richness of DM-haloes in different environments
was made by Gottl\"ober et al. (\cite{gottl03}), using
high-resolution numerical simulations.  Their results show that DM-haloes in
voids have much lower masses than in high-density environments. This
result was recently confirmed by Colberg et al. (\cite{colberg04}).
Their Fig.~14  demonstrates that the most massive  DM-haloes in voids
are about 100 times less massive than the most massive DM-haloes in
general; also the growth of masses of DM-haloes in voids is less rapid
than in general.

\subsection{Interpretation of the environmental dependence of galaxy
  and cluster properties}

One may ask, why are void galaxies and clusters so different from
galaxies and clusters in dense regions?

The evolution of DM-haloes in numerical simulations has been
investigated by a number of authors. Several of these simulations have
been visualized, as an example those by Andrey Kravtsov
({\tt http://cfcp.uchicago.edu/lss/filaments.html}).  In this
simulation the formation of rich clusters and superclusters consisting
of a system of intertwined filaments can be clearly seen.  

We are interested in the difference between the structure of galaxy
systems in high- and low-density environments.  This problem has been
studied in detail by Gottl\"ober et al. (\cite{gottl03}, for earlier
work see references in this paper).  They chose several large
under-dense regions of a diameter of $\sim$20~\Mpc\ (voids defined by bright
simulated galaxies) and re-simulated the evolution of these voids with
a very high mass resolution $4.0\times 10^7~h^{-1} M_{\odot}$. Their
Fig.~2 shows the distribution of dark matter in one of these voids at
the epochs $z=2$ and $z=0$.  At both epochs a well-developed system of
filaments with compact knots (DM-haloes) can be seen.  However, the
masses of these haloes are at both epochs of the order of $10^9~h^{-1}
M_{\odot}$, only the most massive ones have masses of few times of
$10^{10}~h^{-1} M_{\odot}$.  In other words, after the early growth
the knots stop growing.  The volume density of DM-haloes in these voids
is a factor of 10 lower that in the simulation as a whole.

In contrast, the growth of DM-haloes in high-density regions is much
more rapid and continues over the whole period under study.  As shown,
among others, by Frisch et al. (\cite{f95}), the first objects to form
in simulations are rich clusters in superclusters.    Our calculations
presented in the last Section fully confirm these earlier 
conclusions.

\section{Conclusions}

The main results of our study of clusters and superclusters in the SDSS DR1
and the comparison with the results of numerical simulations can be
summarised as follows:

\begin{itemize}
\item{} We have found groups and clusters in the SDSS DR1 data using
  three-dimensional information on the distribution of galaxies.

\item{} Using Gaussian smoothing with the rms scales of 0.8 and 10 \Mpc\ we
  have derived high- and low-resolution luminosity density fields for
  the SDSS DR1 
  data in two equatorial strips; the low-resolution density was used
  as an environmental parameter to describe the large-scale
  environment of groups and clusters.

\item{} New three-dimensional data confirm the environmental dependence
  found earlier using the two-dimensional data: in high-density regions
  (superclusters) groups and clusters are richer and bigger, and
  galaxies themselves are brighter.

\item{} Numerical simulations show a similar tendency: in a high-density
  environment DM-haloes are richer, and have larger bulk motions than
  DM-haloes in a low-density environment.

\item{} Our explanation of the density-luminosity relationship is by
  the combined influence of density perturbations of all scales.
  Superclusters are the regions where the density perturbations
  of large and medium wavelength combine to generate high-density peaks:
  here the overall density is high and the dynamical evolution of
  clusters and galaxies is rapid and continues until the present.
  Voids are regions where large-scale density perturbations have
  negative amplitudes; here, due to medium and small-scale
  perturbations also a filamentary web forms; however, due to the low mean
  density the dynamical evolution is slow and stops at an early epoch.
 
\end{itemize}

\begin{acknowledgements}

The present study was supported by Estonian Science Foundation grant
ETF 4695, and TO 0060058S98.   We are indebted to the SDSS team for their
efforts in carrying the Survey and making its results available to the
astronomical community. P.H. was supported by the Jenny and Antti Wihuri
foundation.

\end{acknowledgements}


\begin{thebibliography}{}

\bibitem[2003]{abazajian03} Abazajian, K., Adelman-McCarthy, J. K.,
	Ag\"ueros, M. A., et al. 2003, AJ, 126, 2081

\bibitem[1958]{abell} Abell, G. 1958, ApJS, 3, 211

\bibitem[1989]{aco} Abell, G., Corwin, H., \& Olowin, R. 1989, ApJS,
70, 1

\bibitem[2004]{balogh04} Balogh, M., Eke, V., Miller, C., et al. 2004,
  MNRAS, 348, 1355, astro-ph/0311379


\bibitem[2000]{bk00} Beisbart, C., \& Kerscher, M. 2000, ApJ, 545, 6

\bibitem[2003]{benn03} Bennett, C. L., Hill, R. S., Hinshaw, G., \etal,
  2003, ApJS, 148, 119

\bibitem[2001]{blanton} Blanton, M. R., Dalcanton, J., Eisenstein,
  D. et al. 2001, AJ, 121, 2358

\bibitem[2004a]{blant04a} Blanton, M.R., Eisenstein, D., Hogg, D.W., et
  al. 2004a, ApJ (accepted), astro-ph/0310453

\bibitem[2004b]{blant04b} Blanton, M.R., Eisenstein, D., Hogg, D.W., et
  al. 2004b, ApJ (submitted), astro-ph/0411037


\bibitem[1998]{bplk98} Bromley, B.J., Press, W.H., Lin, H., et al. 1998,
  ApJ, 505, 25

\bibitem[2004]{colberg04} Colberg, J.M., Sheth, R.K., Diaferio, A.,
  Gao, L., et al. 2004, MNRAS, (in press), astro-ph/0409162 

\bibitem[2004]{cr04} Croton, D.J., Farrar, G. R., Norberg, P. et
  al. 2004, MNRAS (accepted), astro-ph/0407537

\bibitem[1976]{dg76} Davis, M., \& Geller, M.J. 1976, ApJ, 208, 13


\bibitem[2003]{dp03} De Propris, R., Colless, M., Driver, S.P., et
  al. 2003, MNRAS, 342, 725

\bibitem[1980]{dr80} Dressler, A. 1980, ApJ, 236, 351


\bibitem[2003a]{e03a} Einasto, J., Einasto, M., H\"utsi, G., et al.  
2003a,  A\&A, 410, 425, (E03a)

\bibitem[1999]{1999ApJ...519..456E} Einasto, J., Einasto, M., Tago, E., et
  al.  
1999,  ApJ, 519, 456 

\bibitem[2003b]{e03b} Einasto, J.,  H\"utsi, G., Einasto, M., et al.   
2003b,  A\&A, 405, 425, (E03b)

\bibitem[2004a]{e04a} Einasto, J., Tago, E., Einasto, M., \& Saar, E.
  2004a, ''Nearby Large-Scale Structures and the Zone of Avoidance'',
  Cape Town, 28 March - 02 April 2004, eds. A. Fairall \& P. Woudt,
  astro-ph/0408463  (E04a)


\bibitem[1986]{eks86} Einasto, J., Klypin, A.A. \& Saar, E. 1986,
  MNRAS, 219, 457

\bibitem[1991a]{e91a} Einasto, M. 1991a, MNRAS, 250, 802

\bibitem[1991b]{e91b} Einasto, M. 1991b, MNRAS, 252, 261


\bibitem[2003c]{e03c} Einasto, M., Einasto, J., M\"uller, V.,
Hein\"am\"aki, P.,\&  Tucker, D. L. 2003c, A\&A, 401, 851, (E03c)

\bibitem[2003d]{e03d} Einasto, M., Jaaniste, J., Einasto, J., et al.  
2003d, A\&A, 405,  821, (E03d)

\bibitem[2004b]{e04b} Einasto M., Suhhonenko, I., Hein\"am\"aki, P.,
  \& Einasto J. 2004b, A\&A, (in preparation, E04b)


\bibitem[1995]{f95} Frisch, P., Einasto, J., Einasto, M., et al.  
 1995, \aa, 296, 611

\bibitem[2000]{gg00} Girardi, M., \& Giuricin, G. 2000, ApJ, 540, 45

\bibitem[2003]{gottl03} Gottl\"ober, S., Lokas, E. L., Klypin, A., et
  al. 2003, MNRAS, 344, 715 

\bibitem[2003]{hei03} Hein\"am\"aki, P., Einasto, J., Einasto, M., et al. 
2003, A\&A, 397, 63

\bibitem[1982]{hg82} Huchra, J. P., \& Geller, M. J. ApJ, 257, 423


\bibitem[2002]{h02} H\"utsi, G., Einasto, J., Tucker, D.L. et
  al. 2002, astro-ph/0212327

\bibitem[2001]{jfw01} Jenkins, A., Frenk, C.S., White, S.D.M., et al. 2001,
  MNRAS, 321, 372 

\bibitem[1993]{klypin93}  Klypin, A., Holtzman, J., Primack, J., \&
  Reg\"oz, E. 1993, ApJ, 416, 1 

\bibitem[2001]{knebe01} Knebe, A., Green, A. \& Binney, J. 2001,
  MNRAS, 325, 845

\bibitem[1995]{l95} Lindner, U., Einasto, J., Einasto, M., et al.
1995, A\&A, 301, 329

\bibitem[1996]{l96}  Lindner U., {Einasto}, M., {Einasto}, J., et
  al. 1996, A\&A, 314, 1  

\bibitem[1992]{mexd92} Mo, H.J., Einasto, M., Xia, X.Y. \& Deng, Z.G.
  1992, MNRAS, 255, 382

\bibitem[2004]{mo04} Mo, H.J., Yang, X., van den Bosch, F.C., et
  al. 2004, MNRAS, 349, 205

\bibitem[2001]{n01} Norberg, P., Baugh, C.M., Hawkins, E., et
  al. 2001, MNRAS, 328, 64


\bibitem[2002]{n02}  Norberg, P., Baugh, C.M., Hawkins, E., et
  al. 2002, MNRAS, 332, 827

\bibitem[1999]{p99} Peacock, J.P. 1999, Cosmological Physics,
  Cambridge Univ. Press

\bibitem[2001]{p01} Peebles, P.J.E. 2001, ApJ, 557, 495


\bibitem[1974]{ps74} Press, W.H. \& Schechter, P.L. 1974, ApJ, 187, 425


\bibitem[2004]{ragone04} Ragone, C.J., Merch\'ab, M., Muriel, H., et
  al. 2004, MNRAS {in press}, astro-ph/0402155

\bibitem[1976]{S76} Schechter, P. 1976, \apj, 203,
	297

\bibitem[1968]{sersic68} S\'ersic, J.L. 1968, Atlas de Galaxias
  Australes (Cordoba Obs. Astron\'omico)

\bibitem[2002]{s02} Stoughton, C., {Lupton}, R.~H., {Bernardi},
M. et al. 2002, AJ, 123, 485

\bibitem[2004]{t04} Tegmark M., Strauss, M.A., Blanton, M.R., et al.
2004, PhRvD, 69, 103501, astro-ph/0310723


\bibitem[2000]{Tucker00} Tucker, D.L., Oemler, A.Jr., Hashimoto, Y.
et al. 2000, ApJS, 130, 237

\bibitem[1982]{zes82}  Zeldovich, Ya.B., Einasto, J. \&
Shandarin, S.F.   1982, Nature, 300, 407

\bibitem[1961--68]{zwicky} Zwicky, F., Wield, P., Herzog, E.,
Karpowicz, M.  \& Kowal, C.T. 1961--68, Catalogue of Galaxies and
Clusters of Galaxies, 6 volumes.  Pasadena, California Inst. Techn.


\end{thebibliography}
\end{document}